\documentclass[10pt, twocolumn, twoside]{IEEEtran}
\usepackage{amssymb}
\usepackage{amsfonts}

\usepackage{color}

\usepackage{cite}

\ifCLASSINFOpdf
  \usepackage[pdftex]{graphicx}
  \graphicspath{{./Figures/}{./pdf/}{./jpeg/}}
\else
\fi
\usepackage[cmex10]{amsmath}
\usepackage{array}
\usepackage{bbm}                
\usepackage{bbold}              
\usepackage{adjustbox}          
\usepackage{multirow}           
\usepackage{threeparttable}     

\usepackage{stfloats}
\newcommand{\wbar}[1]{\mkern1mu\overline{\mkern-1mu #1\mkern-1mu}\mkern1mu}     
\newcommand{\ubar}[1]{\mkern1mu\underline{\mkern-1mu #1\mkern-1mu}\mkern1mu}    

\begin{document}
%
\title{\vspace{-2mm} \huge Low-Complexity Equalization of MIMO-OSDM}

%
%
%

\author{\vspace{-0.5mm}
        Jing~Han,~\IEEEmembership{Member,~IEEE,}
        Shengqian Ma,
        Yujie Wang,
        and~Geert~Leus,~\IEEEmembership{Fellow,~IEEE}
        \vspace{-6.5mm}}

%
%

\markboth{IEEE TRANSACTIONS ON VEHICULAR TECHNOLOGY}
{Han \MakeLowercase{\textit{et al.}}: Low-Complexity Equalization of MIMO-OSDM}
%



\maketitle

\begin{abstract}
Orthogonal signal-division multiplexing (OSDM) is an attractive alternative to conventional orthogonal frequency-division multiplexing (OFDM) due to its enhanced ability in peak-to-average power ratio (PAPR) reduction. Combining OSDM with multiple-input multiple-output (MIMO) signaling has the potential to achieve high spectral and power efficiency.
However, a direct channel equalization in this case incurs a cubic complexity, which may be expensive for practical use.
To solve the problem, low-complexity per-vector and block equalization algorithms of MIMO-OSDM are proposed in this paper for time-invariant and time-varying channels, respectively.
By exploiting the channel matrix structures, these algorithms have only a linear complexity in the transformed domain.
Simulation results demonstrate their validity and the related performance comparisons.
\end{abstract}




\section{Introduction}
%
%
%
%

Multiple-input multiple-output (MIMO) signaling is a powerful technique to enhance the system spectral efficiency and/or to achieve spatial diversity gain \cite{Paulraj&Gore&Nabar_IEEEJPROC_2004}.
For years there has been a lasting interest on its combination with orthogonal frequency-division multiplexing (OFDM).
One of the main reasons for this is that OFDM is capable of converting a time-invariant (TI) frequency-selective channel into a parallel set of frequency-flat channels, thus enabling low-complexity equalization to mitigate inter-symbol interference (ISI).
However, the high peak-to-average power ratio (PAPR) of OFDM may also lead to a substantial penalty in the
power efficiency of MIMO systems \cite{Bao&Fang&Wan_IEEEJVT_2018}.

As an alternative, orthogonal signal-division multiplexing (OSDM) has recently received much attention, especially in underwater acoustic (UWA) communications \cite{Ebihara&Mizutani_IEEEJOE_2014, Ebihara&Leus_IEEEJOE_2016, Jing&Geert_IEEEJOE_2019, Jing&Geert_IEEETSP_2019}, where the influence of the PAPR is more pronounced.
Compared to conventional OFDM, where the data block is treated as a whole and modulated by a single full-length inverse discrete Fourier transform (IDFT), OSDM has a similar signal structure as vector OFDM in \cite{Yabo&Ngebani&Xiang-Gen_IEEEJSP_2012}, which splits the data block into vectors and performs several component-wise IDFTs among them. By adjusting the vector length, flexible tradeoffs can be achieved between the PAPR and the bandwidth management ability, with two extremes being conventional OFDM and single-carrier block transmission (SCBT) \cite{Yabo&Ngebani&Xiang-Gen_IEEEJSP_2012, Jing&Geert_IEEEJOE_2019}.

Due to this appealing feature, there have been some newly emerging works on MIMO-OSDM.
For example, in \cite{Jing&Geert_IEEECOML_2017} simple Alamouti-like space-time and space-frequency block coding systems were investigated over TI channels, while in \cite{Ebihara&Leus&Ogasawara_Oceans_2018} a spatial multiplexing scheme was designed for time-varying (TV) UWA channels. The latter is particularly attractive for high-rate UWA communications, where the channels are both bandwidth-limited and Doppler-distorted.
However, since the equalizer in \cite{Ebihara&Leus&Ogasawara_Oceans_2018} performs direct channel matrix inversion, it incurs a cubic complexity.
The aim of this paper is to design low-complexity equalizers for MIMO-OSDM by exploiting the channel matrix structure.
Our contributions are as follows.

\begin{itemize}
  \item Over TI channels, only intra-vector ISI exists. As such, a per-vector equalizer is proposed based on the block-diagonal channel matrix structure.
  \item Over TV channels, inter-vector interference (IVI) arises. A block equalizer is proposed based on the block-banded channel matrix approximation by using the complex exponential basis expansion model (CE-BEM).
\end{itemize}
Both equalization algorithms are performed in the transformed domain as their single-input single-output (SISO) counterparts in \cite{Jing&Geert_IEEETSP_2019}; however, they require a judicious (post-)pre-processing of (de)interleaving to enable a linear complexity.

\IEEEpubidadjcol

\textit{Notation}: ${\left(  \cdot  \right)^*}$ stands for conjugate, ${\left(  \cdot  \right)^T}$ for transpose, and ${\left(  \cdot  \right)^H}$ for Hermitian transpose.
We define $[{\bf{x}}] _{m:n}$ as the subvector of $\bf{x}$ from entry $m$ to $n$, and $[{\bf{X}}]_{m:n,p:q}$ as the submatrix of $\bf{X}$ from row $m$ to $n$ and from column $p$ to $q$, where all indices are starting from 0.
Moreover, ${\rm{diag}}\left\{ {\bf{x}} \right\}$ represents a diagonal matrix with $\bf{x}$ on its diagonal, and ${\rm{Diag}}\left\{ {{{\bf{A}}_0}, \ldots ,{{\bf{A}}_{N-1}}} \right\}$ represents a block-diagonal matrix created with the submatrices $\{ {{\bf{A}}_n} \}_{n=0}^{N-1}$.
Also, ${\bf{F}}_N$ stands for the $N \times N$ unitary discrete Fourier transform (DFT) matrix; ${\bf{I}}_N$ and ${\bf{e}}_N (n)$ refer to the $N \times N$ identity matrix and its $n$th column, respectively; ${\bf{0}}_{N}$ denotes the $N \times 1$ all-zero vector;
${\bf{P}}_{M,N}$ is the $MN \times MN$ permutation matrix defined as
\begin{IEEEeqnarray*}{c}
{{\bf{P}}_{M,N}} = \left[ {\begin{array}{*{20}{c}}
{{{\bf{I}}_M} \otimes {\bf{e}}_N^T(0)}\\
{{{\bf{I}}_M} \otimes {\bf{e}}_N^T(1)}\\
 \vdots \\
{{{\bf{I}}_M} \otimes {\bf{e}}_N^T(N - 1)}
\end{array}} \right],
\label{eqn_permut_matrix}
\end{IEEEeqnarray*}
where $\otimes$ denotes the Kronecker product.

\section{Equalization over TI Channels}

We consider a MIMO-OSDM system with $U$ transmitters and $V$ receivers.
The symbol block at the $u$th transmitter is denoted by ${\bf{d}}^{(u)}$ and its length is assumed to be $K=MN$.
Unlike OFDM modulation which uses a single length-$K$ IDFT, OSDM modulation can be expressed as
\begin{IEEEeqnarray}{rCl}
{\bf{s}}^{(u)} = \left( {\bf{F}}_N^H \otimes {\bf{I}}_M \right) {\bf{d}}^{(u)},
\label{eqn_s_u}
\end{IEEEeqnarray}
for $u = 1,2, \ldots, U$. A PAPR reduction is achieved, since it contains $M$ (shorter) IDFTs among $N$ symbol vectors
\begin{IEEEeqnarray}{rCl}
{\bf{d}}_n^{(u)} = {[{\bf{d}}_{}^{(u)}]_{nM:nM + M - 1}},
\label{eqn_dn_u}
\end{IEEEeqnarray}
for $n = 0,1, \ldots, N-1$. Then, after a cyclic prefix (CP) insertion, the block is transmitted through channels.

Let us first assume that all the channels are time invariant and denote the channel impulse response (CIR) between the $u$th transmitter and the $v$th receiver by ${\bf{c}}^{(v,u)} = [ {c_0^{(v,u)},c_1^{(v,u)}, \ldots ,c_L^{(v,u)}} ]^T$, where $L$ is the channel order.
In this case, the received signal block at the $v$th receiver after CP removal has the form
\begin{IEEEeqnarray}{rCl}
{\bf{r}}^{(v)} = \sum\limits_{u = 1}^{U} {{\bf{\tilde C}}^{(v,u)} {\bf{s}}^{(u)}}  + {{\bf{w}}^{(v)}},
\label{eqn_r_v}
\end{IEEEeqnarray}
for $v = 1, 2, \ldots, V$, where ${\bf{\tilde C}}^{(v,u)}$ is the $K \times K$ circulant channel matrix with its first column being $[ {\bf{c}}^{(v,u)T}, {\bf{0}}_{K - L - 1}^T ]^T$; ${\bf{w}}^{(v)}$ is the $K \times 1$ received noise.

Subsequently, OSDM demodulation is performed as
\begin{IEEEeqnarray}{rCl}
{{\bf{x}}^{(v)}} = \left( {{\bf{F}}_N^{} \otimes {\bf{I}}_M^{}} \right){{\bf{r}}^{(v)}} = \sum\limits_{u = 1}^U {{{\bf{C}}^{(v,u)}}{{\bf{d}}^{(u)}}}  + {{\bf{z}}^{(v)}},
\label{eqn_x_v}
\end{IEEEeqnarray}
where ${{\bf{C}}^{(v,u)}} = ( {{\bf{F}}_N^{} \otimes {\bf{I}}_M^{}} ){{\bf{\tilde C}}^{(v,u)}} ( {{\bf{F}}_N^H \otimes {\bf{I}}_M^{}})$ is termed the composite channel matrix; ${\bf{z}}^{(v)}$ is the $K \times 1$ demodulated noise.
It can be easily verified from (\ref{eqn_s_u}) and (\ref{eqn_x_v}) that, when $M = 1$ ($N = K$) and $M = K$ ($N = 1$), the signal model of MIMO-OSDM is equivalent to that of MIMO-OFDM and MIMO-SCBT, respectively. In this sense, MIMO-OSDM can be deemed as a more generalized MIMO scheme.

\begin{figure}[!t]
\vskip -2mm
\centering
\includegraphics[width=2.52in]{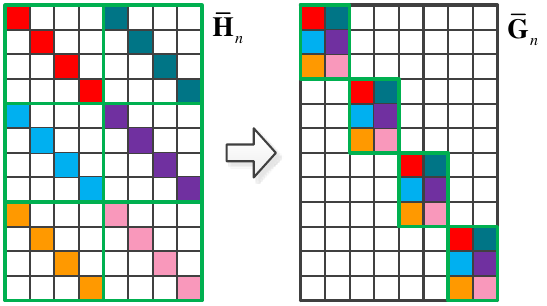}       
\vskip -2mm
\caption{An example of the TI channel matrix structures of ${\bf{\wbar H}}_n$ and ${\bf{\wbar G}}_n$ with $U=2$, $V=3$ and $M=4$.}
\vskip -2mm
\label{Figure_1}
\end{figure}

Moreover, for TI channels, it is known that the composite channel matrix has the block-diagonal structure \cite{Jing&Geert_IEEETSP_2019}
\begin{IEEEeqnarray}{rCl}
{\bf{C}}^{(v,u)} = {\rm{Diag}}\left\{ {{\bf{H}}_{\rm{0}}^{(v,u)},{\bf{H}}_{\rm{1}}^{(v,u)}, \ldots ,{\bf{H}}_{N - 1}^{(v,u)}} \right\},
\label{eqn_C_vu}
\end{IEEEeqnarray}
where
\begin{IEEEeqnarray}{rCl}
{\bf{H}}_n^{(v,u)} & = & {\bf{\Lambda }}_M^{nH}{\bf{F}}_M^H{\bf{\wbar H}}_n^{(v,u)}{\bf{F}}_M^{}{\bf{\Lambda }}_M^n,
\label{eqn_Hn_vu}     \\
{\bf{\wbar H}}_n^{(v,u)} & = & {\rm{diag}} \{ {H_n^{(v,u)},H_{n + N}^{(v,u)}, \ldots ,H_{n + \left( {M - 1} \right)N}^{(v,u)}}  \},
\label{eqn_Hbarn_vu}
\end{IEEEeqnarray}
${\bf{\Lambda }}_M^n = {\rm{diag}} \{ {[ {1,{e^{ - j\frac{{2\pi n}}{K}}}, \ldots ,{e^{ - j\frac{{2\pi n}}{K}\left( {M - 1} \right)}}} ]^T} \}$ and $H_k^{(v,u)} = \sum\nolimits_{l = 0}^L {c_l^{(v,u)}{e^{ - j\frac{{2\pi }}{K}lk}}}$.
Given this matrix structure, we partition ${\bf{x}}^{(v)}$ and ${\bf{z}}^{(v)}$ into $N$ vectors of length $M$ as in (\ref{eqn_dn_u}). Since only intra-vector ISI exists in this case, the channel equalization can be decoupled on each vector. Specifically, by defining ${\bf{x}}_n^{(v)} = {[{\bf{x}}_{}^{(v)}]_{nM:nM + M - 1}}$ and ${\bf{z}}_n^{(v)} = {[{\bf{z}}_{}^{(v)}]_{nM:nM + M - 1}}$ for $n = 0,1, \ldots, N-1$ as the $n$th demodulated vector and noise vector, respectively, it can be readily obtained that
\begin{IEEEeqnarray}{rCl}
{{\bf{x}}_n} = {{\bf{H}}_n}{{\bf{d}}_n} + {{\bf{z}}_n},
\label{eqn_x_n}
\end{IEEEeqnarray}
where we have stacked all the $n$th vectors, i.e, ${\bf{d}}_n = [{\bf{d}}_n^{(1)T},{\bf{d}}_n^{(2)T}, \ldots ,{\bf{d}}_n^{(U)T}]^T$, ${\bf{x}}_n = [{\bf{x}}_n^{(1)T},{\bf{x}}_n^{(2)T}, \ldots ,{\bf{x}}_n^{(V)T}]^T$, ${\bf{z}}_n = [{\bf{z}}_n^{(1)T},{\bf{z}}_n^{(2)T}, \ldots ,{\bf{z}}_n^{(V)T}]^T$, and
\begin{IEEEeqnarray}{rCl}
{{\bf{H}}_n} =
\begin{bmatrix}
   {{\bf{H}}_n^{(1,1)}} & {{\bf{H}}_n^{(1,2)}} &  \ldots  & {{\bf{H}}_n^{(1,U)}}  \\
   {{\bf{H}}_n^{(2,1)}} & {{\bf{H}}_n^{(2,2)}} & {\ldots} & {{\bf{H}}_n^{(2,U)}}  \\
   {\vdots} & {\vdots} & {\ddots} & {\vdots}  \\
   {{\bf{H}}_n^{(V,1)}} & {{\bf{H}}_n^{(V,2)}} & {\ldots} & {{\bf{H}}_n^{(V,U)}}
\end{bmatrix}.
\label{eqn_H_n}
\end{IEEEeqnarray}

Throughout this paper, we assume that the input symbols on all transmitters are independent and identically distributed (i.i.d.) with unit power, while the noise samples on different receivers are zero mean with the same variance ${\sigma}^2$. Therefore, based on (\ref{eqn_x_n}), the minimum mean-square error (MMSE) equalization algorithm can be written as
\begin{IEEEeqnarray}{rCl}
{{\bf{\hat d}}_n} = \left( {\bf{R}}_n^{-1} {\bf{H}}_n^H \right) {{\bf{x}}_n}.
\label{eqn_dnhat_direct}
\end{IEEEeqnarray}
where ${\bf{R}}_n = {\bf{H}}_n^H{{\bf{H}}_n} + {\sigma ^2}{{\bf{I}}_{UM}}$ is an $UM \times UM$ matrix. Since a straightforward computation of ${\bf{R}}_n^{-1}$ will incur a complexity of ${\mathcal{O}}(U^3 M^3)$, to ease the computational burden, we substitute (\ref{eqn_Hn_vu}) into (\ref{eqn_H_n}), yielding
\begin{IEEEeqnarray}{rCl}
{\bf{H}}_n = {\bf{\Phi }}_{n,V}^H {{\bf{\wbar H}}_n^{}} {{\bf{\Phi }}_{n,U}^{}},
\label{eqn_Hn_fact}
\end{IEEEeqnarray}
where ${{\bf{\Phi }}_{n,i}} = {{\bf{I}}_i} \otimes ({\bf{F}}_M^{}{\bf{\Lambda }}_M^n)$; ${{\bf{\wbar H}}_n}$ has a similar structure as ${\bf{H}}_n$ in (\ref{eqn_H_n}) with its blocks replaced by the diagonal matrices $\{ {\bf{\wbar H}}_n^{(v,u)} \}$.
Furthermore, as illustrated in Fig.~1, ${{\bf{\wbar H}}_n}$ can actually be interleaved into a block-diagonal matrix, i.e.,
\begin{IEEEeqnarray}{rCl}
{\bf{\wbar G}}_n
& = & {{\bf{P}}_{V,M}} {{\bf{\wbar H}}_n} {\bf{P}}_{U,M}^H
\nonumber \\
& = & {\rm{Diag}}\left\{ {{\bf{\wbar G}}_{n,0}},{\bf{\wbar G}}_{n,0}, \ldots ,{\bf{\wbar G}}_{n,M-1} \right\},
\label{eqn_G_n}
\end{IEEEeqnarray}
with ${\bf{\wbar G}}_{n,m}$ being blocks of size $V \times U$.

\begin{figure}[!t]
\vskip -2mm
\centering
\includegraphics[width=3.5in]{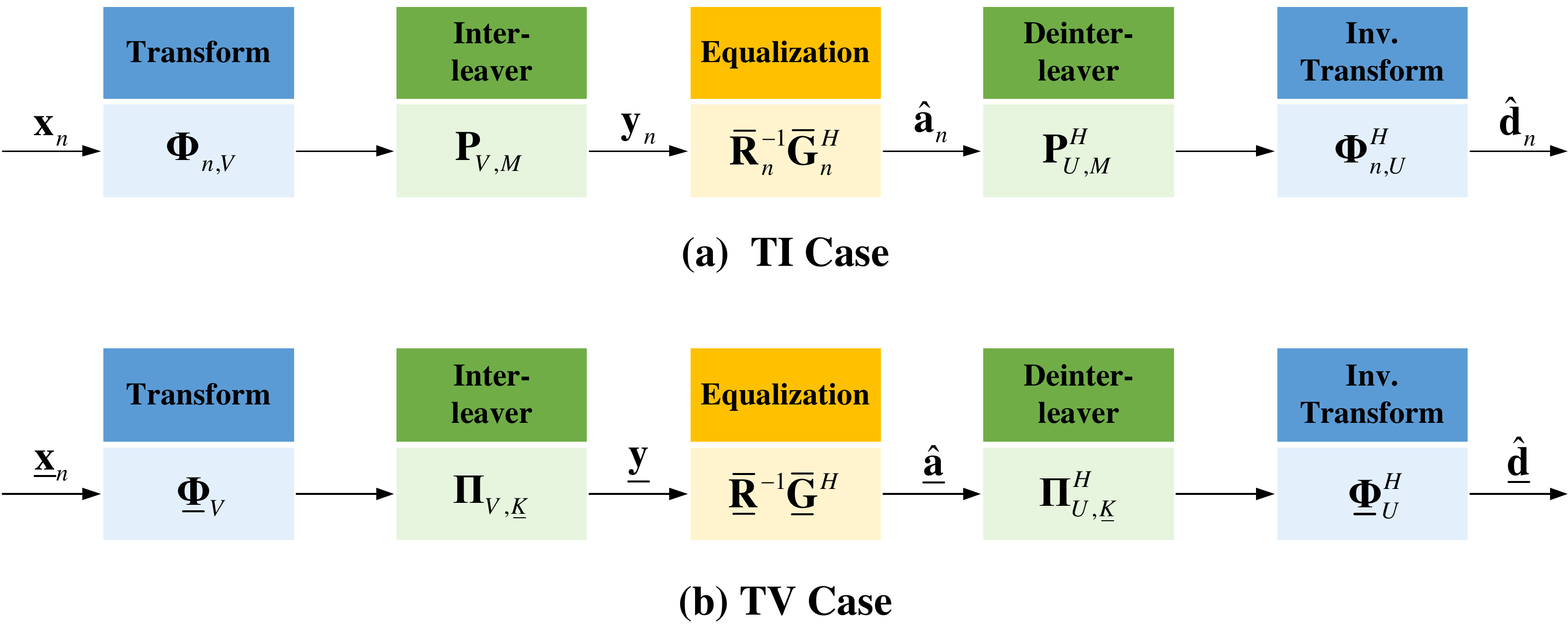}       
\vskip -2mm
\caption{Proposed TI and TV channel equalization schemes for MIMO-OSDM.}
\vskip -2mm
\label{Figure_2}
\end{figure}

\begin{figure*}[!t]
\vskip -3mm
\centering
\includegraphics[width=6.5in]{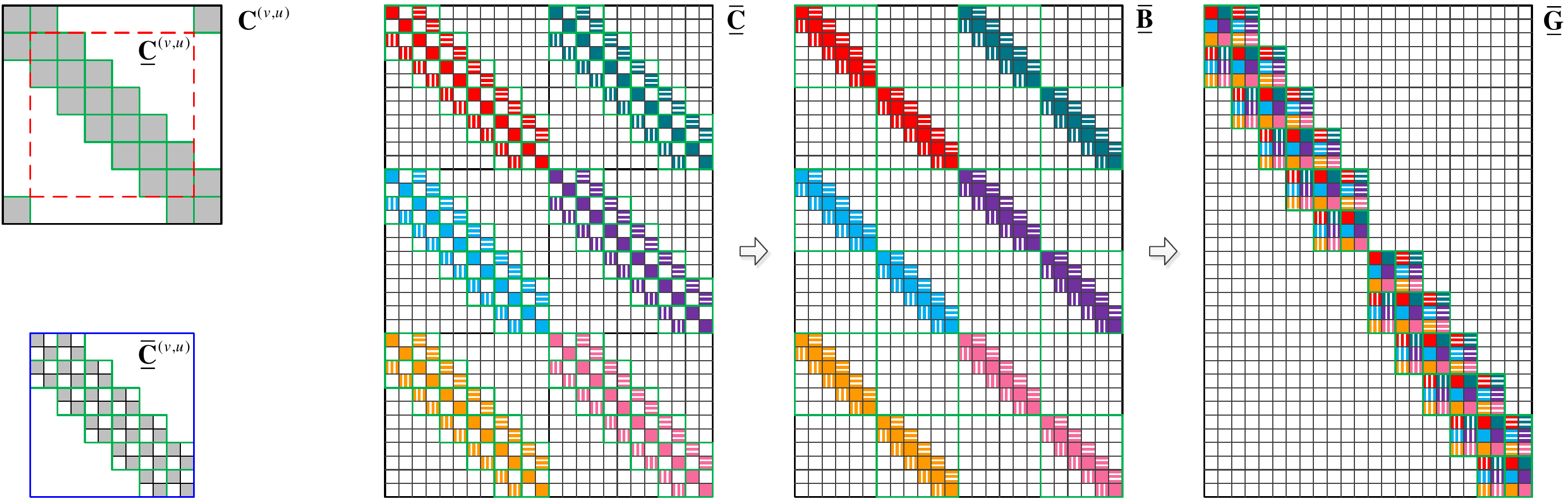}       
\vskip -2mm
\caption{An example of the TV channel matrix structures with $U=2$, $V=3$, $M=2$, $N=8$ and $Q=1$.}
\vskip -3mm
\label{Figure_3}
\end{figure*}

Based on the matrix factorizations in (\ref{eqn_Hn_fact}) and (\ref{eqn_G_n}), the symbol estimation in (\ref{eqn_dnhat_direct}) can be rewritten as
\begin{IEEEeqnarray}{rCl}
{\bf{\hat d}}_n  =  {\bf{\Phi }}_{n,U}^H {\bf{P}}_{U,M}^H \left( {\bf{\wbar R}}_n^{-1} {\bf{\wbar G}}_n^H \right) {\bf{P}}_{V,M}^{}{\bf{\Phi }}_{n,V}^{}{{\bf{x}}_n},
\label{eqn_dnhat_LC}
\end{IEEEeqnarray}
where ${\bf{\wbar R}}_n = {\bf{\wbar G}}_n^H {\bf{\wbar G}}_n + {\sigma ^2}{\bf{I}}_{UM}$.
As shown in Fig.~2(a), (\ref{eqn_dnhat_LC}) actually corresponds to a low-complexity implementation of MIMO-OSDM equalization over TI channels, which consists of five steps:

\begin{enumerate}
  \item Transform the demodulated vector ${{\bf{x}}_n}$ by ${\bf{\Phi }}_{n,V}$.
  \item Interleave the transformed vector by ${\bf{P}}_{V,M}$.
  \item Equalize the interleaved vector (denoted by ${\bf{y}}_n$) to obtain the transformed symbol estimate (denoted by ${\bf{\hat a}}_n$), i.e.,
        \begin{IEEEeqnarray}{rCl}
        {{\bf{\hat a}}_n} = \left( {\bf{\wbar R}}_n^{-1} {\bf{\wbar G}}_n^H \right) {{\bf{y}}_n}.
        \label{eqn_an_hat}
        \end{IEEEeqnarray}
  \item Deinterleave the output of the equalizer by ${\bf{P}}_{U,M}^H$.
  \item Perform the inverse transform ${\bf{\Phi }}_{n,U}^H$ to finally obtain ${\bf{\hat d}}_n$.
\end{enumerate}

In this implementation, steps 1 and 5 involve $V$ DFTs and $U$ IDFTs of length $M$, respectively, resulting in a complexity of $\mathcal{O}((U+V)M{\log _2}M)$.
As for step 3, the transformed-domain equalization in (\ref{eqn_an_hat}) may look similar to that in (\ref{eqn_dnhat_direct}); however, its computation is more tractable. To be specific, it can be seen that ${\bf{\wbar R}}_n$ is a block-diagonal matrix with $M$ blocks of size $U \times U$ on its diagonal, so the inversion in (\ref{eqn_an_hat}) has only a complexity of $\mathcal{O} (U^3 M)$, which is linear in the vector length $M$.
Given the fact that the values of $U$ and $V$ are typically not large, the total complexity of (\ref{eqn_dnhat_LC}) will be easy to handle.

\section{Equalization over TV Channels}

We proceed to consider the equalization of MIMO-OSDM over TV channels.
The CIR between the $u$th transmitter and the $v$th receiver is now denoted by $\{ c_{k,l}^{(v,u)} \}$, where the index $k$ is added to embody the time dependence of the CIR.
And in this case $\{ {\bf{C}}^{v,u} \}$  no longer have the block-diagonal structure as in (\ref{eqn_C_vu}); instead, they are generally full matrices. As a result, IVI arises in OSDM, which is a counterpart of inter-carrier interference (ICI) in OFDM.

For simplicity, the CE-BEM in \cite{Ebihara&Leus_IEEEJOE_2016, Jing&Geert_IEEETSP_2019, Ebihara&Leus&Ogasawara_Oceans_2018} is adopted to approximate the TV CIR. It utilizes complex exponential bases to capture the channel time variations within each block, i.e.,
\begin{IEEEeqnarray}{rCl}
c_{k,l}^{(v,u)} = \sum\limits_{q=-Q}^Q {h_{q,l}^{(v,u)}{e^{j\frac{2\pi}{K}qk}}},
\label{eqn_CIR_TV}
\end{IEEEeqnarray}
for $k = 0, 1, \ldots, K-1$, where $Q$ is the discrete Doppler spread and $\{ h_{q,l}^{(v,u)} \}$ are the BEM coefficients. With this model, the number of channel parameters on each delay tap $l$ is reduced from $K$ to $2Q+1$.
Moreover, based on (\ref{eqn_CIR_TV}), we have the TV channel matrix
\begin{IEEEeqnarray}{rCl}
{\bf{\tilde C}}^{(v,u)} = \sum\limits_{q=-Q}^Q {{\bf{\tilde \Gamma }}_K^q{\bf{\tilde C}}_q^{(v,u)}},
\label{eqn_Ctilde_vu_TV}
\end{IEEEeqnarray}
where ${\bf{\tilde \Gamma}}_K^q = {\rm{diag}} \{ [ {1,{e^{j\frac{{2\pi }}{K}q}}, \ldots ,{e^{j\frac{{2\pi }}{K}q\left( {K - 1} \right)}}} ]^T \}$; ${\bf{\tilde C}}_q^{(v,u)}$ is a circulant matrix with its first column equal to ${\bf{h}}_q^{(v,u)} = [ h_{q,0}^{(v,u)}, h_{q,1}^{(v,u)}, \ldots , h_{q,L}^{(v,u)} ]^T$ appended by $K - L - 1$ zeros.
Note that, as shown in the left of Fig.~3, the composite channel matrix ${\bf{C}}^{(v,u)}$ corresponding to (\ref{eqn_Ctilde_vu_TV}) is (cyclically) block-banded with block semi-bandwidth (BSB) $Q$ (see \cite[Proposition~3]{Jing&Geert_IEEETSP_2019} for a proof), which lays the foundation for our low-complexity equalization algorithm in this section.

Specifically, the presence of IVI excludes the use of the per-vector equalization algorithm previously designed for TI channels. We thus consider a block equalization for TV channels, which jointly estimates all symbol vectors in an OSDM block.
Moreover, to achieve a low-complexity implementation, we again resort to matrix factorization of the blocks in ${\bf{C}}^{(v,u)}$.
Let ${{\bf{C}}_{n,n'}^{(v,u)}} = {[ {\bf{C}}^{(v,u)} ]_{nM:nM + M - 1,n'M:n'M + M - 1}}$ be the $(n,n')$th block of ${\bf{C}}^{(v,u)}$. It has been shown in \cite[Proposition~4]{Jing&Geert_IEEETSP_2019} that only blocks in the main band of ${\bf{C}}^{(v,u)}$ can be diagonalized. More specifically, only when $\left| {n - n'} \right| \le Q$, we have
\begin{IEEEeqnarray}{rCl}
{\bf{C}}_{n,n'}^{(v,u)} = {\bf{\Lambda }}_M^{nH}{\bf{F}}_M^H{{\bf{\wbar H}}_{n-n', n'}^{(v,u)}}{\bf{F}}_M^{}{\bf{\Lambda }}_M^{n'},
\label{eqn_C_uv_blk_fact}
\end{IEEEeqnarray}
where
\begin{IEEEeqnarray}{rCl}
{{\bf{\wbar H}}}_{q,n}^{(v,u)} = {\rm{diag}} \{ [ {H_{q,n}^{(v,u)}},{H_{q,N + n}^{(v,u)}}, \ldots ,{H_{q, (M-1)N+n}^{(v,u)}} ]^T \},
\IEEEeqnarraynumspace
\label{eqn_Hbar_qn}
\end{IEEEeqnarray}
and ${H_{q,k}^{(v,u)}} = \sum\nolimits_{l = 0}^L {{h_{q,l}^{(v,u)}}{e^{ - j\frac{{2\pi }}{K}lk}}}$ for $k = 0, 1, \ldots K-1$.

To eliminate the blocks in the bottom-left and top-right corners of ${\bf{C}}^{(v,u)}$ (which cannot be diagonalized), at each transmitter we place $Q$ zero vectors at both edges of the symbol block, i.e., ${\bf{d}}^{(u)} = [{\bf{0}}_{MQ}^T, {\bf{\ubar d}}^{(u)T}, {\bf{0}}_{MQ}^T ]^T$, where ${\bf{\ubar d}}^{(u)} = {\bf{T}} {\bf{d}}^{(u)}$ contains the middle ${\ubar N} = N - 2Q$ payload vectors with ${\bf{T}} = {\left[ {{{\bf{I}}_K}} \right]_{QM:\left( {N - Q} \right)M - 1, 1:K}}$. Accordingly, at each receiver the demodulated block is truncated as ${\bf{\ubar x}}^{(v)} = {\bf{T}} {\bf{x}}^{(v)}$. Then, it can be obtained that
\begin{IEEEeqnarray}{rCl}
{\bf{\ubar x}}^{(v)} = \sum\limits_{u = 1}^U {{{\bf{\ubar C}}^{(v,u)}}{{\bf{\ubar d}}^{(u)}}}  + {\bf{\ubar z}}^{(v)},
\label{eqn_xubar_v}
\end{IEEEeqnarray}
where ${\bf{\ubar C}}^{(v,u)} = {\bf{T}} {\bf{C}}^{(v,u)} {\bf{T}}^H$ and ${\bf{\ubar z}}^{(v)}$ is the noise term. As shown in Fig.~3, ${\bf{\ubar C}}^{(v,u)}$ is a standard (not cyclically) block-banded matrix. Based on (\ref{eqn_C_uv_blk_fact}), it can be further factorized into
\begin{IEEEeqnarray}{rCl}
{\bf{\ubar C}}^{(v,u)} = {\bf{\ubar \Omega}}^H {\bf{\wbar{\ubar C}}}^{(v,u)} {\bf{\ubar \Omega}},
\label{eqn_Cubar_uv_fact}
\end{IEEEeqnarray}
where ${\bf{\ubar \Omega}} = {\rm{Diag}} \{ {\bf{F}}_M^{}{\bf{\Lambda }}_M^Q, \, {\bf{F}}_M^{}{\bf{\Lambda }}_M^{Q + 1}, \ldots , \, {\bf{F}}_M^{}{\bf{\Lambda }}_M^{N - Q - 1} \}$; ${\bf{\wbar{\ubar C}}}^{(v,u)}$ has the same matrix structure as ${\bf{\ubar C}}^{(v,u)}$, but with all its nonzero blocks being diagonal (see Fig.~3).

Now, let us stack all these blocks of length ${\ubar K} = M {\ubar N}$, and define ${\bf{\ubar d}} = [{\bf{\ubar d}}^{(1)T},{\bf{\ubar d}}^{(2)T}, \ldots ,{\bf{\ubar d}}^{(U)T}]^T$, ${\bf{\ubar x}} = [{\bf{\ubar x}}^{(1)T},{\bf{\ubar x}}^{(2)T}, \ldots ,{\bf{\ubar x}}^{(V)T}]^T$, ${\bf{\ubar z}} = [{\bf{\ubar z}}^{(1)T},{\bf{\ubar z}}^{(2)T}, \ldots ,{\bf{\ubar z}}^{(V)T}]^T$.
From (\ref{eqn_xubar_v}) and (\ref{eqn_Cubar_uv_fact}), we then have the signal model
\begin{IEEEeqnarray}{rCl}
{\bf{\ubar x}} = {\bf{\ubar C}} {\bf{\ubar d}} + {\bf{\ubar z}},
\label{eqn_xubar}
\end{IEEEeqnarray}
where ${\bf{\ubar C}} = {{\bf{\ubar \Phi}}_V^H} {{\bf{\wbar{\ubar C}}}} {{\bf{\ubar \Phi}}_U}$, with ${{\bf{\ubar \Phi}}_i} = {\bf{I}}_i \otimes {\bf{\ubar \Omega}}$ and
\begin{IEEEeqnarray}{rCl}
{\bf{\wbar{\ubar C}}} =
\begin{bmatrix}
   {\bf{\wbar{\ubar C}}}^{(1,1)} & {\bf{\wbar{\ubar C}}}^{(1,2)} &  \ldots  & {\bf{\wbar{\ubar C}}}^{(1,U)}  \\
   {\bf{\wbar{\ubar C}}}^{(2,1)} & {\bf{\wbar{\ubar C}}}^{(2,2)} & {\ldots} & {\bf{\wbar{\ubar C}}}^{(2,U)}  \\
   {\vdots} & {\vdots} & {\ddots} & {\vdots}  \\
   {\bf{\wbar{\ubar C}}}^{(V,1)} & {\bf{\wbar{\ubar C}}}^{(V,2)} & {\ldots} & {\bf{\wbar{\ubar C}}}^{(V,U)}
\end{bmatrix}.
\label{eqn_Cbar2}
\end{IEEEeqnarray}
As illustrated in the right of Fig.~3, the matrix structure of $\bf{\wbar{\ubar C}}$ can be further simplified by interleaving, i.e.,
\begin{IEEEeqnarray}{rCl}
{\bf{\wbar{\ubar G}}}
= {{\bf{P}}_{V,{\ubar K}}} {\bf{\wbar{\ubar B}}} {\bf{P}}_{U,{\ubar K}}^H
= {{\bf{\Pi}}_{V,{\ubar K}}} {\bf{\wbar{\ubar C}}} {\bf{\Pi}}_{U,{\ubar K}}^H,
\label{eqn_Cbar2_Intlv}
\end{IEEEeqnarray}
where ${\bf{\Pi}}_{i,{\ubar K}} = {{\bf{P}}_{i,{\ubar K}}} ( {{{\bf{I}}_i} \otimes {{\bf{P}}_{{\ubar N}, M}}} )$, and the resulting matrix ${\bf{\wbar{\ubar G}}}$ is block-banded with block size $V \times U$ and BSB $Q$.

Therefore, the MMSE equalization of MIMO-OSDM over TV channels can be written as
\begin{IEEEeqnarray}{rCl}
{\bf{\hat{\ubar d}}}
& = & \left( {\bf{\ubar R}}^{-1} {\bf{\ubar C}}^H \right) {\bf{\ubar x}}
\label{eqn_dubar_direct}\\
& = & {\bf{\ubar \Phi}}_U^H{\bf{\Pi }}_{U,{\ubar K}}^H\left( {{{\bf{\wbar{\ubar R}}}^{ - 1}}{{\bf{\wbar{\ubar G}}}^H}} \right){\bf{\Pi }}_{V,{\ubar K}}^{}{\bf{\ubar \Phi}}_V^{}{\bf{\ubar x}},
\label{eqn_dubar_LC}
\end{IEEEeqnarray}
where ${\bf{\ubar R}} = {{\bf{\ubar C}}^H}{\bf{\ubar C}} + {\sigma ^2}{{\bf{I}}_{U{\ubar K}}}$ and ${\bf{\wbar{\ubar R}}} = {{\bf{\wbar{\ubar G}}}^H}{\bf{\wbar{\ubar G}}} + {\sigma ^2}{{\bf{I}}_{U{\ubar K}}}$.
The above equations (\ref{eqn_dubar_direct}) and (\ref{eqn_dubar_LC}) represent the direct and low-complexity implementations, respectively.
While (\ref{eqn_dubar_direct}) suffers from a cubic complexity of $\mathcal{O} \{ U^3 {\ubar K}^3 \}$, as shown in Fig.~2(b), (\ref{eqn_dubar_LC}) actually takes the same strategy as (\ref{eqn_dnhat_LC}) to reduce the computational burden.
Specifically, here ${\bf{\ubar x}}$ is first transformed and interleaved into ${\bf{\ubar y}} = {\bf{\Pi }}_{V,{\ubar K}}^{}{\bf{\ubar \Phi}}_V^{}{\bf{\ubar x}}$, on which the TV channel equalization is then performed as
\begin{IEEEeqnarray}{rCl}
{\bf{\hat{\ubar a}}} = ( {{\bf{\wbar{\ubar R}}}^{ - 1}} {{\bf{\wbar{\ubar G}}}^H} ) {\bf{\ubar y}},
\label{eqn_aubar_hat}
\end{IEEEeqnarray}
and finally the estimate of the symbol blocks is produced by ${\bf{\hat{\ubar d}}} = {\bf{\ubar \Phi}}_U^H{\bf{\Pi }}_{U,{\ubar K}}^H {\bf{\hat{\ubar a}}}$.
It is easy to see that $\bf{\wbar{\ubar R}}$ is a block-banded matrix  with block size $U \times U$ and BSB $2Q$.
As a result, we can use the block LDL$^H$ algorithm in \cite{Jing&Geert_IEEETSP_2019} to compute the matrix inversion in (\ref{eqn_aubar_hat}), and the complexity is only $\mathcal{O} \{ U^3 Q^2 {\ubar K} \}$.

\section{Numerical Results}

In this section, the bit-error rate (BER) performances of the proposed equalization algorithms are evaluated by numerical simulations.
We here consider a MIMO-OSDM system in a UWA communication scenario.
At each transmitter, OSDM blocks are composed of $K = 1024$ quaternary phase-shift keying (QPSK) symbols with symbol period $T_{\rm{s}} = 0.25$~ms, and thus the block duration is $T = K T_{\rm{s}}= 256$~ms.
The MIMO channel is assumed to have an order of $L = 24$, corresponding to a multipath delay spread of ${\tau}_{\max} = LT_{\rm s} = 6$~ms, with all the taps Rayleigh distributed and generated from a uniform power delay profile.
We first focus on TI channel equalization in Fig.~4, where $2 \times 2$, $2 \times 3$ and $2 \times 4$ transmissions are investigated and the vector length is set to $M = 1$ (i.e., MIMO-OFDM), $4$ and $16$. As expected, a better system performance is achieved with a larger $V$ due to the increased spatial diversity. Meanwhile, it is seen that the BER also improves as the vector length gets longer. This can be attributed to the intra-vector frequency diversity specific in OSDM systems \cite{Yabo&Ngebani&Xiang-Gen_IEEEJSP_2012, Jing&Geert_IEEEJOE_2019}.

In Fig.~5, we further introduce the channel time variation and simulate it by a U-shaped Doppler spectrum with normalized Doppler spread $f_d T = 0.25$ and $0.5$. Here, the $2 \times 3$ MIMO configuration is adopted and the OSDM vector length is fixed to $M = 16$. It can be seen that, compared to the direct equalization algorithm in (\ref{eqn_dubar_direct}) using the full channel matrix, the proposed low-complexity equalizer in (\ref{eqn_dubar_LC}) leads to a error floor due to its block-banded channel matrix approximation based on the CE-BEM.
However, their performance gap can be narrowed by increasing $Q$. Moreover, a significant reduction in complexity can be achieved. As an example, when $M = 16$ and $Q = 4$,  the complexity of the proposed equalizer is only $0.008\%$ of that of the direct equalization method.

\begin{figure}[!t]
\centering
\includegraphics[width=2.5in]{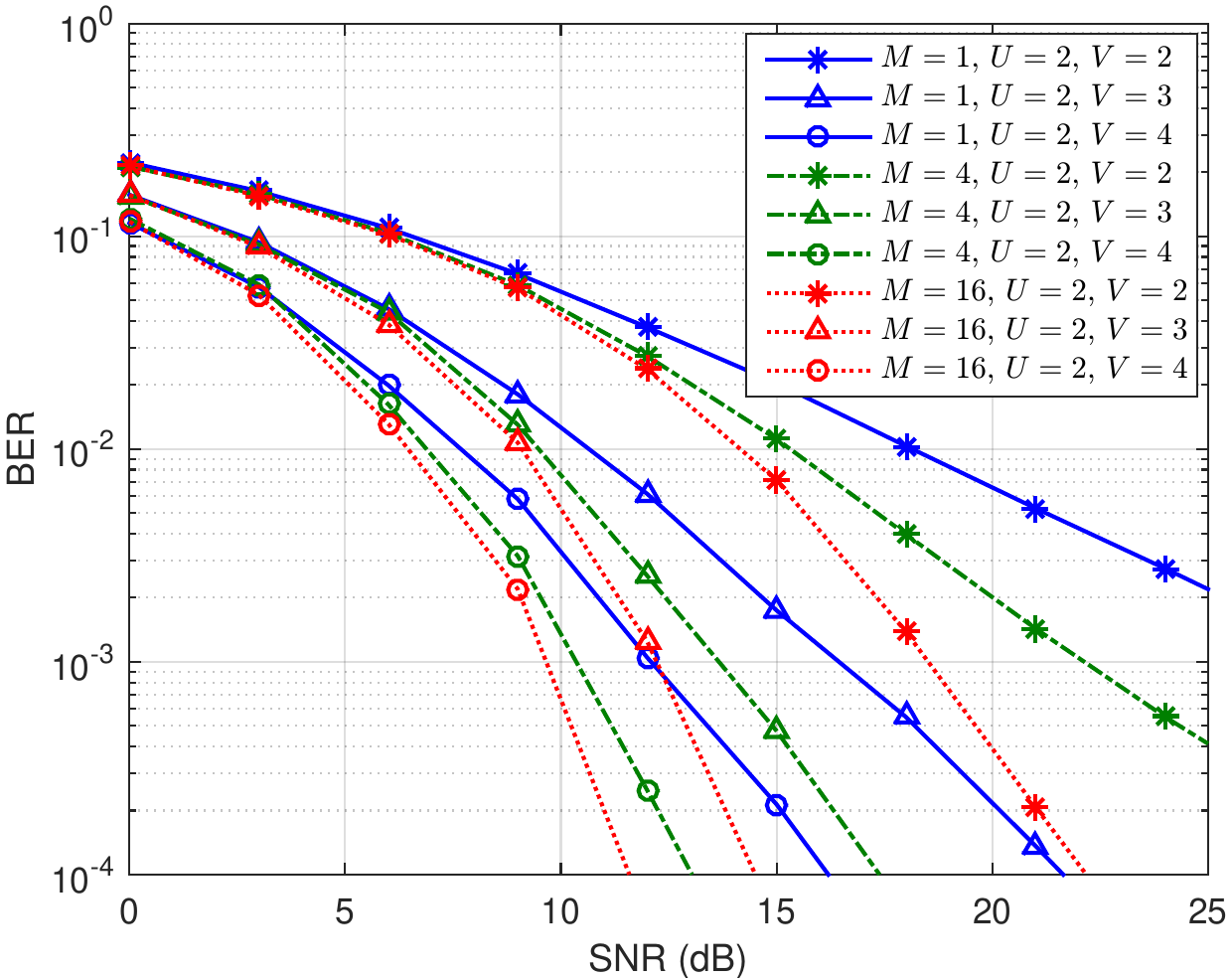}
\vskip -0.7em
\caption{BER performance of MIMO-OSDM equalization over TI channels.}
\label{fig3}
\vskip -0.7em
\end{figure}

\begin{figure}[!t]
\centering
\includegraphics[width=2.5in]{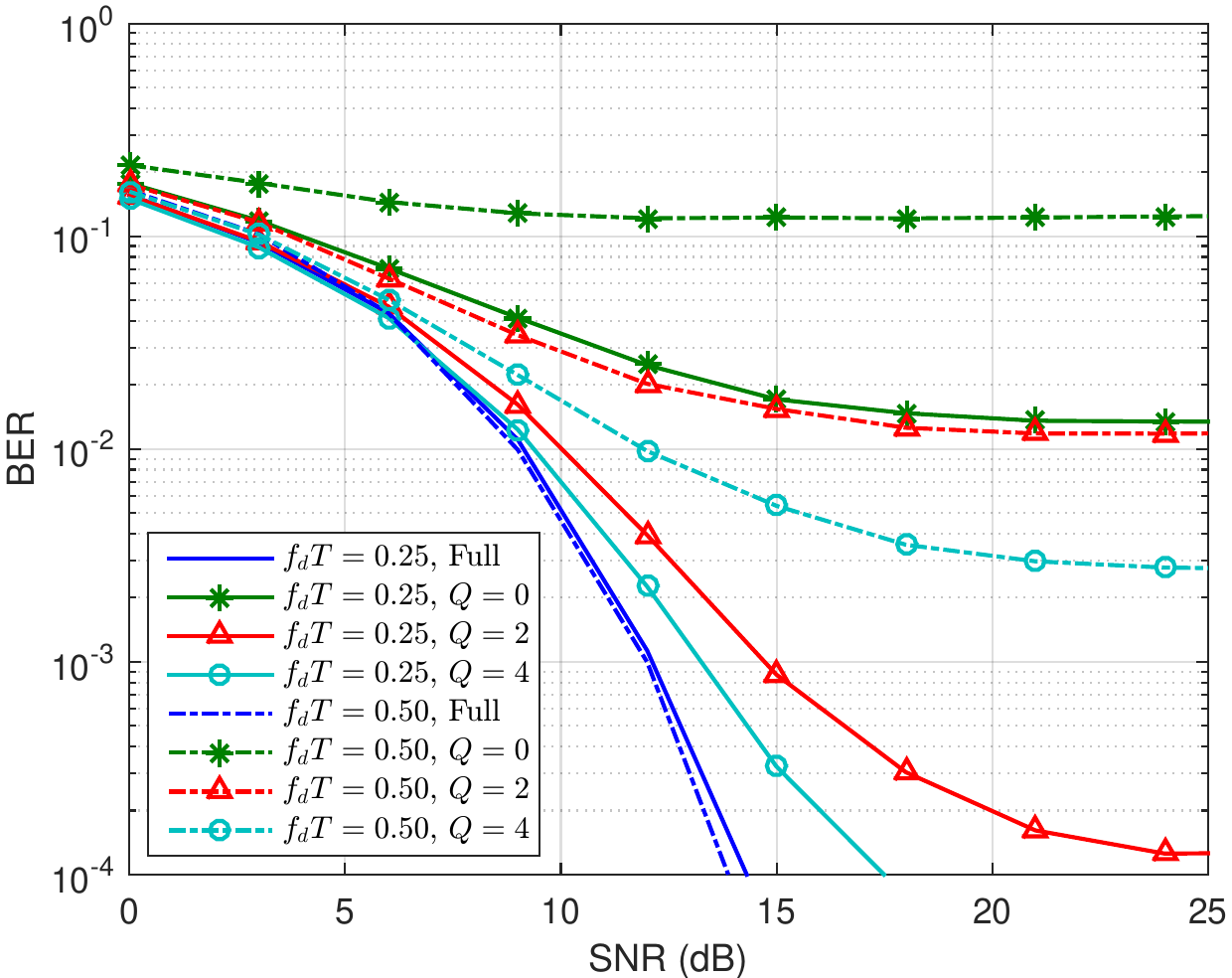}
\vskip -0.7em
\caption{BER performance of MIMO-OSDM equalization over TV channels.}
\label{fig4}
\vskip -0.7em
\end{figure}

\section{Conclusions}

Low-complexity equalization algorithms of MIMO-OSDM are proposed in this paper for TI and TV channels (Fig. 2). Compared to the direct equalization method of cubic complexity, they have only a linear complexity in the transformed domain and thus are promising for practical use.

\appendices
%


\ifCLASSOPTIONcaptionsoff
  \newpage
\fi



\bibliographystyle{IEEEtran}
\bibliography{IEEEabrv,../bib/HJ}    

%
%

%





\end{document}